\documentclass[aps,prc,12pt,showpacs,showkeys,superscriptaddress,groupedaddress]{revtex4-1}

\usepackage[english]{babel}
\usepackage{indentfirst}
\usepackage{amssymb}
\usepackage{braket}
\usepackage{bm}
\usepackage{amsxtra}
\usepackage{amsmath}
\usepackage{supertabular}
\usepackage{multirow}
\usepackage{color}
\usepackage [mathcal]{eucal}
\usepackage{graphicx}
\usepackage{graphics}
\usepackage{lipsum}

\def\n4lo{$\mathrm{N}^4\mathrm{LO}$}
\def\n3lo{$\mathrm{N}^3\mathrm{LO}$}

     \def\chiral4lo{$\mathrm{N}^4\mathrm{LO}$}

\begin{document}

\noindent
\title{Proton-Nucleus Elastic Scattering: Comparison between Phenomenological and Microscopic Optical Potentials
}

\author{Matteo Vorabbi$^{1}$}
\author{Paolo Finelli$^{2}$}
\author{Carlotta Giusti$^{3}$}

\affiliation{$~^{1}$TRIUMF, 4004 Wesbrook Mall, Vancouver, British Columbia, V6T 2A3, Canada
}

\affiliation{$~^{2}$Dipartimento di Fisica e Astronomia, 
Universit\`{a} degli Studi di Bologna and \\
INFN, Sezione di Bologna, Via Irnerio 46, I-40126 Bologna, Italy
}

\affiliation{$~^{3}$Dipartimento di Fisica,  
Universit\`a degli Studi di Pavia and \\
INFN, Sezione di Pavia,  Via A. Bassi 6, I-27100 Pavia, Italy
}

\date{\today}


\begin{abstract} 

{\bf Background:} Elastic scattering is a very important process to understand nuclear interactions in finite nuclei. Despite decades of efforts, the goal of reaching a coherent
description of this physical process in terms of microscopic forces is still far from being completed.

{\bf Purpose:} In previous papers~\cite{Vorabbi:2015nra,Vorabbi:2017rvk} we derived a nonrelativistic theoretical optical potential from nucleon-nucleon chiral potentials at
fourth (N$^3$LO) and fifth order (N$^4$LO). We checked convergence patterns and established theoretical error bands. With this work we study the performances of our
optical potential in comparison with those of a successful nonrelativistic phenomenological optical potential in the description of elastic proton scattering data on several isotopic
chains at energies around and above 200 MeV. 

{\bf Methods:} We use the same framework and the same approximations as in Refs.~\cite{Vorabbi:2015nra,Vorabbi:2017rvk}, where the nonrelativistic optical potential is derived
at the first-order term within the spectator expansion of the multiple scattering theory and adopting the impulse approximation and the optimum factorization approximation. 

{\bf Results:} The cross sections and analyzing powers for elastic proton scattering off calcium, nickel, tin, and lead isotopes are presented for several incident proton energies, exploring the range $156 \le E \le 333$ MeV, where experimental data are available. In addition, we provide theoretical predictions for $^{56}$Ni at 400 MeV, which is of interest for the future experiments at EXL.  

{\bf Conclusions:} Our results indicate that microscopic optical potentials derived from nucleon-nucleon chiral potentials at N$^4$LO can provide reliable predictions for the cross section and the analyzing power both of stable and exotic nuclei, even at energies where the reliability of the chiral expansion starts to be questionable.

\end{abstract}


\maketitle


\section{Introduction}
\label{sec_intro}

The scattering process of an incident nucleon off a target nucleus is a widespread experimental tool for investigating, with specific nuclear reactions, the  different properties of a
nuclear system. Elastic scattering is probably the main event occurring in the nucleon-nucleus (NA) scattering and measurements of cross sections and polarization observables
in elastic proton-nucleus (pA) scattering have provided a lot of detailed information on nuclear properties \cite{PaetzSchieck,Glendenning}.

A huge amount of experimental data has been collected over the last years concerning stable nuclei (usually with proton or neutron numbers corresponding to some
magic configurations) but nowadays one of the most active areas of research in nuclear physics is to understand the properties of nuclei far from the beta-stability line.  
A number of radioactive ion beam facilities will be used in next years for this purpose. In particular, we would like to mention the FAIR project, with the section dedicated to
electromagnetic and light hadronic probes (EXL) \cite{fair, exl}, where the structure of unstable exotic nuclei in light-ion scattering experiments at intermediate energies will be
extensively studied.
Some preliminary measurements have already been performed by investigating the reaction $^{56}$Ni(p,p)$^{56}$Ni at an energy of $400$ MeV in inverse kinematics \cite{refId0}.
The authors of Ref. \cite{refId0} claim that the preliminary results are very promising and demonstrate the feasibility of the intended program of EXL. This result strongly supports
the need of a reliable description of the interaction of a nucleon with stable and unstable nuclei. Unfortunately, such processes are characterized by many-body effects that make
their theoretical description an extremely hard task.

A very useful framework to achieve this goal is provided by the theoretical concept of the Optical Potential (OP), where the complicated nature of the NA interaction is described introducing a
complex effective potential whose real part describes the average interaction between the projectile and the target, and the imaginary part the effect of all inelastic processes
which tend to deplete the flux in the elastic channel \cite{FESHBACH1958357}. 
The OP was originally employed to analyze the NA elastic scattering data, but its use has been afterwards extended to inelastic scattering and to a wide variety of nuclear reactions.

Different OPs for elastic NA scattering have been derived either by phenomenological analyses of experimental data or by a more fundamental microscopic
calculation. Phenomenological OPs are obtained assuming an analytical form of the potential which depends on some free parameters specifying the well and the geometry of the
system and that are determined by a fitting procedure over a set of available experimental data of elastic pA scattering. This approach provides good OPs, which perform very well in
many regions of the nuclear chart and for several energy ranges where data are available, but which may lack predictive power when applied to situations where experimental data are
not yet available. 
Instead, microscopic OPs are derived starting from the Nucleon-Nucleon (NN) interaction; they can be obtained using different NN potentials and different methods
depending on the mass of the target and on the energy of the reaction of interest, and do not contain free adjustable parameters. A recent list of the different approaches can be
found in Ref.~\cite{dickhoff}. Being the result of a model and not of a fitting procedure, microscopic OPs should have more theoretical content and might have a more general
predictive power than phenomenological OPs, but the approximations which are needed to reduce the complexity of the original many-body problem, whose exact solution is for
complex nuclei beyond our present capabilities, might give a poorer agreement with available empirical data.

In Ref.~\cite{Vorabbi:2015nra} we constructed a microscopic OP for elastic pA scattering starting from NN chiral potentials derived up to N$^3$LO in the chiral expansion and we studied the chiral convergence of the NN potential in reproducing the pA scattering observables. The OP was obtained at the first-order term within the spectator expansion of the nonrelativistic multiple scattering theory and adopting the impulse approximation and the optimum factorization approximation.
In a subsequent work~\cite{Vorabbi:2017rvk} we adopted the same model to obtain the OP and we studied the chiral convergence of a new generation of NN chiral interactions derived up to N$^4$LO. Our conclusion was that
the convergence has been reached at N$^4$LO. 

In this work we perform a systematic investigation of the predictive power of our microscopic OP derived in Ref.~\cite{Vorabbi:2017rvk} from different chiral potentials at N$^4$LO and
of the successful phenomenological OP of Refs. \cite{Koning:2003zz,KONING2014187} in comparison with available data for the observables of elastic proton scattering on different
isotopic chains, located in different areas of the nuclear chart. Results are presented for several proton energies around and above 200 MeV, with the aim to test the upper energy
limit of applicability of our OP before the chiral expansion scheme breaks down.

The paper is organized as follows: In Sec.~\ref{sec_model} we summarize the main features of both microscopic and phenomenological approaches to the OP. In
Sec.~\ref{sec_res} we show and discuss the results for the observables of elastic pA scattering. Finally, in Sec~\ref{sec_concl} we draw our conclusions.


\section{Theoretical Models}
\label{sec_model}

The underlying assumption on which the OP is based is that the interaction between the projectile
and the target nucleus can be modelled by a complex mean field potential. The numerical solution of the Schr\"odinger 
equation (or the Lippman-Schwinger equation in the momentum space representation) 
with this complex potential brings valuable information, i.e. the differential cross section $d\sigma/d\Omega$, the analyzing power $A_y$, and the spin rotation $Q$ 
among the others \cite{PaetzSchieck,Glendenning}. 

In our approach we limit ourselves to the study of the elastic channel, defined as the case in which the target and the projectile remain in the stationary state, 
at energies where a non-relativistic description can be conveniently applied. Given a state $|{\bf k}_{el}^\dagger \rangle$ which describes the relative motion of 
both collision partners, after some manipulations, a Schr\"odinger equation can be derived for the projectile and the target in the relative motion that reads as follows
\begin{equation}
\label{opt1}
\left( T + V_{opt} - E \right) |{\bf k}_{el}^\dagger \rangle= 0 \; . 
\end{equation}
$T$ is the kinetic energy operator, $E$ is the available energy in the elastic channel, and $V_{opt}$ is the so-called generalized OP \cite{FESHBACH1958357} 
defined as
\begin{equation}
\label{opt2}
V_{opt} = V_{2b} + V_{2b} Q \frac{1}{E - QHQ + i \eta}Q V_{2b} \; ,
\end{equation}
where $V_{2b}$ is a two-body potential,  and $P$ and $Q$ are idempotent projection operators introduced to isolate the contribution of the elastic channel: $P$ projects onto the elastic channel and $Q$ onto the complementary space, i.e. onto all the non-elastic channels, for which the following relations hold: $Q = 1 - P$, and $QP = PQ= 0$. 

The first term in Eq.~(\ref{opt2}) is the contribution of the static two-body interaction, while the second term takes into account the effect of non-elastic channels in the space $Q$. 
This term is the dynamic part of the OP and depends on the energy of the elastic channel. As a consequence, the Feshbach approach naturally leads to an 
energy-dependent OP $V_{opt}(E)$.

The general structure of $V_{opt}$ is extremely complicated and can be simplified for some specific applications. 
We refer the reader to Refs. \cite{PaetzSchieck,Glendenning,Jeukenne:1976uy,Mahaux:1985zz} for exhaustive discussions about the approximations which are necessary to deal 
with the treatment of $V_{opt}$. Here we will restrict our considerations to the bare essentials.

Generally speaking, there are two available methods for the construction of an OP.

In the microscopic approach, one starts from a realistic NN interaction (i.e. able to reproduce the experimental NN phase shifts with a 
$\chi^2$ per datum very close to one \cite{Machleidt:1989tm,Machleidt:2011zz,Epelbaum:2008ga}) and an educated guess for the radial density of the 
target \cite{PaetzSchieck,Glendenning,Jeukenne:1976uy}. A suitable combination of these two terms (a procedure usually called "folding") produces the optical potential $V_{opt}$.
The main features of the microscopic OP are the independence from phenomenological inputs and, 
in particular with the most recent NN microscopic potentials, the ability to assess reliable error estimates (see Ref. \cite{Epelbaum:2014efa} for extensive discussions about this topic).
In the ideal case where no approximations are made to derive the microscopic OP, the absence of phenomenology would lead to accurate predictions that are probably better than
those obtained with a phenomenological OP, but in practice the full calculation of the OP turns out to be too complicated and the approximations that must necessarily be introduced reduce the accuracy and the reliability of these predictions.

On the other hand, a more pragmatic phenomenological approach can be pursued with the adoption of an analytical form of the potential, i.e. like a Woods-Saxon shape, 
where the adjustable parameters are fitted to a set of available experimental data \cite{Varner:1991zz}.

Since a lot of efforts has been put over the last years on both methods, we believe that it can be useful
to make a comparison between the above mentioned approaches. For this purpose, we decided to use  our recent microscopic OPs derived \cite{Vorabbi:2017rvk} from NN chiral interactions at N$^4$LO \cite{Epelbaum:2014sza,Epelbaum:2014efa,Entem:2014msa,Entem:2017gor} and the most recent analysis by 
Koning {\it et al.} \cite{Koning:2003zz} who developed a very successful nonrelativistic phenomenological OP (KD) for energies below 200 MeV but also with an extension up to 1 GeV \cite{KONING2014187}. 


\subsection{Microscopic optical potentials at \chiral4lo}
\label{sec_chiral}

The theoretical justification for the description of the NA optical potential in terms of the microscopical NN interaction has 
been addressed for the first time by Watson {\it et al.} \cite{Riesenfeld:1956zza} and then formalized by Kerman {\it et al.} (KMT) \cite{Kerman:1959fr}, 
where the so-called multiple scattering approach to the NA optical potential is expressed by a series expansion of the free NN scattering amplitudes.
Over the last decades several authors made important contributions to this approach. Just to mention the most relevant ones, we would like to remind the works with the KMT optimum
factorized OP \cite{PhysRevC.30.1861,PhysRevC.40.881}, the calculation of the full-folding OP with harmonic oscillator
densities \cite{PhysRevC.41.814,PhysRevLett.63.605,PhysRevC.41.2188,PhysRevC.41.2257}, the calculation of the second-order OP in the multiple scattering
theory \cite{PhysRevC.46.279}, the calculation of the medium contributions to the first-order OP \cite{PhysRevC.48.2956,PhysRevC.51.1418,PhysRevC.52.1992}, and
the calculation of the full-folding OP with realistic densities \cite{PhysRevC.56.2080}.
Concerning the inclusion of medium effects we also want to mention the works based on the $g$ matrix of Amos {\it et al.} \cite{Amos2002} and
Arellano {\it et al.} \cite{PhysRevC.52.301,PhysRevC.84.034606}.

In Refs. \cite{Vorabbi:2015nra,Vorabbi:2017rvk} a microscopic OP was obtained at the first-order term within the spectator expansion of the nonrelativistic multiple scattering theory,
corresponding to the single-scattering approximation. The impulse approximation was adopted, where nuclear binding forces on the interacting target nucleon are neglected, as
well as the optimum factorization approximation, where the two basic ingredients of the calculations, {\it i.e.} the nuclear density and the NN $t$ matrix, are factorized.
We refer the reader to Refs. \cite{Vorabbi:2015nra,Vorabbi:2017rvk} for all relevant details and an exhaustive bibliography.
In the momentum space, the factorized $V_{opt}$ is obtained as 
\begin{equation}
V_{opt} ({\bf q}, {\bf K}; \omega) \sim \sum_{N=n,p} t_{pN} \left[
{\bf q},\frac{A+1}{A} {\bf K}; \omega
\right] \rho_N (q) \; ,
\end{equation} 
where $t_{pN}$ represents the proton-proton (pp)
and proton-neutron (pn) free $t$ matrix evaluated at a fixed energy $\omega$, $\rho_N$ the neutron and
proton profile density, and the momentum variables ${\bf k}$ and ${\bf k}^\prime$ are conveniently expressed by the variables ${\bf q} \equiv {\bf k}^\prime - {\bf k}$ and
${\bf K} \equiv \frac{1}{2} ({\bf k}^\prime + {\bf k})$ (see Sect. II of Ref. \cite{Vorabbi:2015nra}). 

For the neutron and proton densities of the target nucleus we use as in Refs. \cite{Vorabbi:2015nra,Vorabbi:2017rvk} a Relativistic Mean-Field (RMF)
description \cite{Niksic20141808}, which has been quite successful in the description of ground state and excited state properties of finite nuclei, in particular in a Density
Dependent Meson Exchange (DDME) version, where the couplings between mesonic and baryonic fields are assumed as functions of the density itself \cite{PhysRevC.66.024306}. 
We are aware that a phenomenological description of the target is not fully consistent with the goal of a microscopic description of elastic NA scattering. In a very recent paper
\cite{Gennari:2017yez} a microscopic OP has been derived using {\it ab initio} translationally invariant nonlocal one-body nuclear densities computed within the No-Core Shell Model
(NCSM) approach \cite{BARRETT2013131}, which is a technique particularly well suited for the description of light nuclei. Indeed the use of a nonlocal {\it ab initio} density improves
significantly the agreement with data of elastic proton scattering off $^4$He and $^{12}$C, while for $^{16}$O no significant improvement is obtained in comparison with the RMF results.
The work reported in Ref. \cite{Gennari:2017yez} represents a great leap forward towards the construction of a microscopic OP for light nuclei, but the aim of our present work is to
investigate the predictive power of a microscopic OP over a wide range of nuclei and isotopic chains in different regions of the nuclear chart.

For the NN interaction we use here two different versions of the chiral potentials at fifth order (N$^4$LO) recently derived by 
Epelbaum, Krebs, and Mei\ss ner (EKM) \cite{Epelbaum:2014sza,Epelbaum:2014efa} and Entem, Machleidt, and Nosyk (EMN) \cite{Entem:2014msa,Entem:2017gor}. 
As explained in Ref. \cite{Vorabbi:2017rvk}, the two versions of the chiral N$^4$LO potentials have significant differences concerning the renormalization procedures and we
follow the same prescriptions adopted there. The strategy followed for the EKM potentials \cite{Epelbaum:2014sza,Epelbaum:2014efa} consists in a coordinate space regularization 
for the long-range contributions $V_{\rm long} ({\bm r}) $, by the introduction of $f\left(\frac{r}{R} \right) = \left(1 -\exp\left( -\frac{r^2}{R^2}\right) \right)^n$,
and a conventional momentum space regularization for the contact  (short-range) terms, with a cutoff $\Lambda = 2R^{-1}$. Five choices of $R$ are available: $0.8, 0.9, 1.0, 1.1$,
and $1.2$ fm, leading to five different potentials.
 
On the other hand, for the EMN potentials, a slightly more conventional approach was pursued  \cite{Entem:2014msa, Entem:2017gor}. 
A spectral function regularization, with a cutoff $\tilde{\Lambda} \simeq 700$ MeV, was employed to regularize the loop contributions and a conventional regulator function,
with $\Lambda = 450,500$, and $550$ MeV, to deal with divergences in the Lippman-Schwinger equation. For all details we 
refer the reader to Refs. \cite{Entem:2014msa, Entem:2017gor,Vorabbi:2017rvk}.

The aim of the present work is to test the predictive power of our microscopic OP in comparison with available experimental data and it can be useful to show the uncertainties
on the predictions produced by NN chiral potentials obtained with different values of the regularization parameters.
For this purpose, all calculations have been performed with three of the EKM \cite{Epelbaum:2014sza,Epelbaum:2014efa} potentials, corresponding to $R=0.8, 0.9$,
and $1.0$ fm, and with two of the EMN \cite{Entem:2014msa, Entem:2017gor} potentials, corresponding to $\Lambda = 500$  and $550$ MeV.
In all the figures presented in Sec.~\ref{sec_res} the bands give the differences produced by changing $R$ for EKM (red bands) and $\Lambda$ for EMN (green bands).
Thus the bands have here a different meaning than in Ref. \cite{Vorabbi:2017rvk}, where the EKM and EMN NN chiral potentials at N$^4$LO were also used.
The aim of Ref. \cite{Vorabbi:2017rvk} was to investigate the convergence and to assess the theoretical errors associated with the truncation of the chiral expansion and the
bands were given to investigate these issues.
We also showed in Ref. \cite{Vorabbi:2017rvk} that EKM calculations based on different values of $R$ are quite close and consistent with each other 
(although, as remarked in Ref. \cite{Epelbaum:2014sza}, larger values of R are probably less accurate due to a larger influence of cutoff artifacts).
The same assumption can be made about the EMN potentials: changing the cutoffs does not lead to sizeable differences in the $\chi^2/$datum 
(see Tab.VIII in Ref. \cite{Entem:2017gor}) and it is safe to perform calculations with only two potentials.
Because we want to explore elastic scattering at energies around and above 200 MeV, we exclude the EKM potentials with $R = 1.1$ and $1.2$ fm and the EMN potential with 
$\Lambda = 450$ MeV. We are confident that for our present purposes showing results with only a limited set of NN chiral potentials will not affect our conclusions in any way.


\subsection{Phenomenological potentials}
\label{sec_talys}

One of the most recent and successful phenomenological OP has been developed by Koning {\it et al.} \cite{Koning:2003zz}.
As quoted in the original paper, the authors provided a phenomenological OP able to challenge the best microscopic approaches 
in terms of predictive power. 

The phenomenological OP $V_{opt}$ can be separated into a real ($V_R$) and an imaginary ($V_I$) part. Both contributions can be expressed as follows
\begin{equation}
V_{opt,i}  = - V_{C,i} (r, E) - V_{S,i}  (r, E) + V_{LS,i} (r, E)~{\bf l} \cdot  {\bf s} \quad ,\quad i=\{ R,I\} \; ,  
\end{equation}
in terms of central ($V_{C}$), surface dependent ($V_{S}$), and spin-orbit ($V_{LS}$) components.
All the components can be separated in energy-dependent well depths and energy-independent shape functions as $V(r, E) \sim \tilde{V}(E)f(r)$, 
where the radial functions $f$ usually resemble a Wood-Saxon shape. 

The potential of Ref. \cite{Koning:2003zz} is a so-called "global" OP, which means that the free adjustable parameters are fitted for
a wide range of nuclei ($24 \le A \le 249$) and of incident energies ($1$ keV $\le E \le 200$ MeV) with some parametric dependence of the coefficients in terms of the target
mass number $A$ and of the incident energy $E$. 
An alternative choice, not adopted in Ref. \cite{Koning:2003zz}, would be to produce an OP for each single target nucleus.
We refer the reader to Ref. \cite{Koning:2003zz} for more details.
Recently, an extension of the OP of Ref. \cite{Koning:2003zz} up to 1 GeV has been proposed \cite{KONING2014187}. It is generally believed that above $\sim 180$ MeV the
Schr\"odinger picture of the phenomenological OP should be taken over by a Dirac approach \cite{talys}, but the extension was done just with the aim to test at which energy the validity of the
predictions of the nonrelativistic OP fail. 
We are aware that above 200 MeV an approach based on the Dirac equation would probably be a more consistent choice, but since we are interested in testing the limit of applicability
of our (nonrelativistic) microscopic OP we will use such an extension to perform some benchmark calculations at center-of-mass energies close to 300 MeV.
All the calculations have been performed by ECIS-06 \cite{ecis} as a subroutine in the TALYS software \cite{talys, talys2}.


\section{Results}
\label{sec_res}

The aim of the present paper is to investigate and compare the predictive power of our microscopic OP derived from the 
EKM~\cite{Epelbaum:2014sza,Epelbaum:2014efa} and EMN~\cite{Entem:2014msa, Entem:2017gor} chiral potentials at N$^4$LO 
and of the phenomenological global OP KD derived by 
Koning {\it et al.} \cite{Koning:2003zz,KONING2014187} in comparison with available data of elastic pA scattering. 
To this aim, in this section we present and discuss the predictions of the different  OPs for the differential cross 
section ${\frac{{\rm d}\sigma}{{\rm d}\Omega}}$, presented as ratio to the Rutherford cross 
section, ${\frac{{\rm d}\sigma}{{\rm d}\Omega}}/{\frac{{\rm d}\sigma}{{{\rm d}\Omega}}}_{\rm Ruth}$, 
and analyzing power $A_y$ of proton elastic scattering over a wide range of nuclei and isotopes chains, from oxygen to lead, 
and for proton energies between 156 and 333 MeV, for which experimental data are available.

The energy range considered for our investigation was chosen on the basis of the assumptions and approximations adopted 
in the derivation of the theoretical OP. In particular, the impulse approximation does not allow us to use our microscopic 
OP with enough confidence at much lower energies, where we can expect that the phenomenological KD potential is able to give 
a better agreement with the experimental data. The upper energy limit is determined by the fact that the EKM and EMN chiral 
potentials are able to describe NN scattering observables up to 300 MeV \cite{Epelbaum:2014sza,Epelbaum:2014efa,Entem:2014msa,Entem:2017gor}. 
The phenomenological global KD potential was originally constructed for energies up to 200 MeV \cite{Koning:2003zz} and it was 
then extended up to 1 GeV \cite{KONING2014187}.
It can therefore be interesting to test and compare the validity of the predictions of both microscopic and phenomenological OPs up to about 300 MeV. 

\begin{figure}[t]
\begin{center}
\includegraphics[scale=0.65]{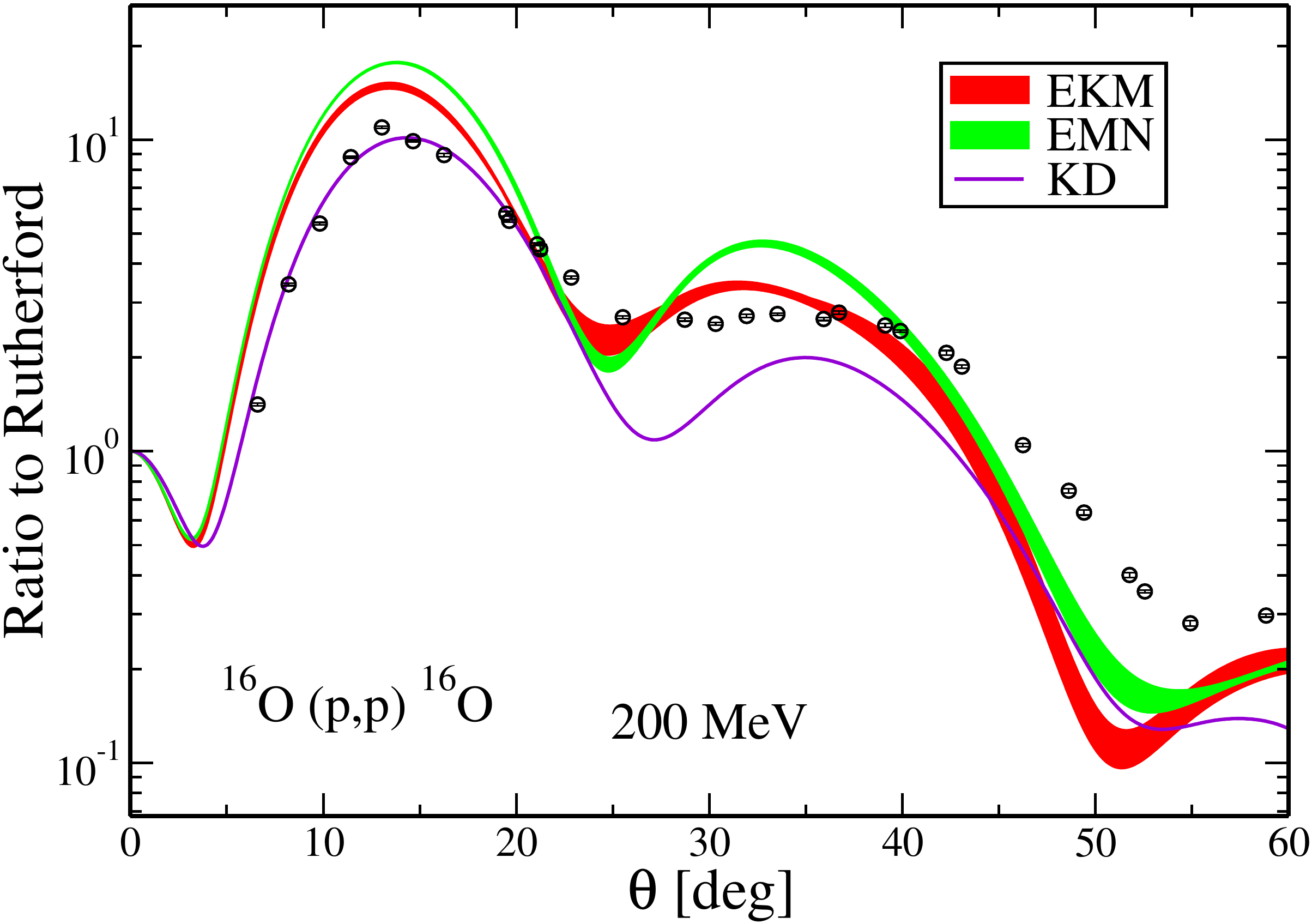}
\caption{ (Color online) Ratio of the differential cross section to the Rutherford cross section as a function of 
the center-of-mass scattering angle $\theta$ for elastic proton scattering off ${}^{16}$O.  Calculations are 
performed at $E= 200$ MeV (laboratory energy) with the microscopic OPs derived from the
EKM~\cite{Epelbaum:2014sza,Epelbaum:2014efa}(EKM, red band)  and EMN~\cite{Entem:2014msa, Entem:2017gor} 
(EMN, green band) NN chiral potentials at N$^4$LO and with the phenomenological global OP of Ref.~\cite{talys} 
(KD, violet line). The interpretation of the bands is explained in the text. Experimental data from Refs. \cite{kelly,exfor}.
\label{fig_o} }
\end{center}
\end{figure}

In Ref. \cite{Vorabbi:2017rvk} we compared the results obtained with different versions of EKM and EMN chiral potentials at 
$\mathrm{N}^4\mathrm{LO}$ for the pp and pn Wolfenstein amplitudes and for the scattering observables of elastic proton scattering 
from $^{16}$O, $^{12}$C, and $^{40}$Ca nuclei  at an incident proton energy $E= 200$ MeV. For the sake of comparison with our 
previous work, we show in Fig.~\ref{fig_o} the ratio of the differential cross section to the Rutherford cross section for 
elastic proton scattering off  $^{16}$O at $E = 200$  MeV. The results obtained with the EKM and EMN potentials and with the KD optical potential are compared
with the experimental data taken from Refs. \cite{kelly,exfor}.
The EKM and EMN results correspond to the results  shown in Fig.~2 of Ref. \cite{Vorabbi:2017rvk} for the differential cross 
section (of course with a different meaning of the bands) and give a reasonable, although not perfect, agreement with data. 
The experimental ratio is slightly overestimated at lower angles and somewhat underestimated  for $\theta \ge 50^{\circ}$. 
The differences between the EKM and EMN results are small and not crucial, EKM gives a smaller cross section around the maxima 
and therefore a somewhat better agreement with the data in this region. The bands, representing the uncertainties on the regularization 
of the NN chiral potentials, are generally small and not influential for the comparison with data.
The KD result gives a good description of the experimental cross section for $\theta \le 20^{\circ}$ and underpredicts the data for larger angles.  
We point out, to be honest, that KD was obtained for nuclei in the mass range $24 \le A \le 209$ while $^{16}$O is  
below this range. We present the result only for the sake of comparison. 

\begin{figure}[t]
\begin{center}
\includegraphics[scale=0.6]{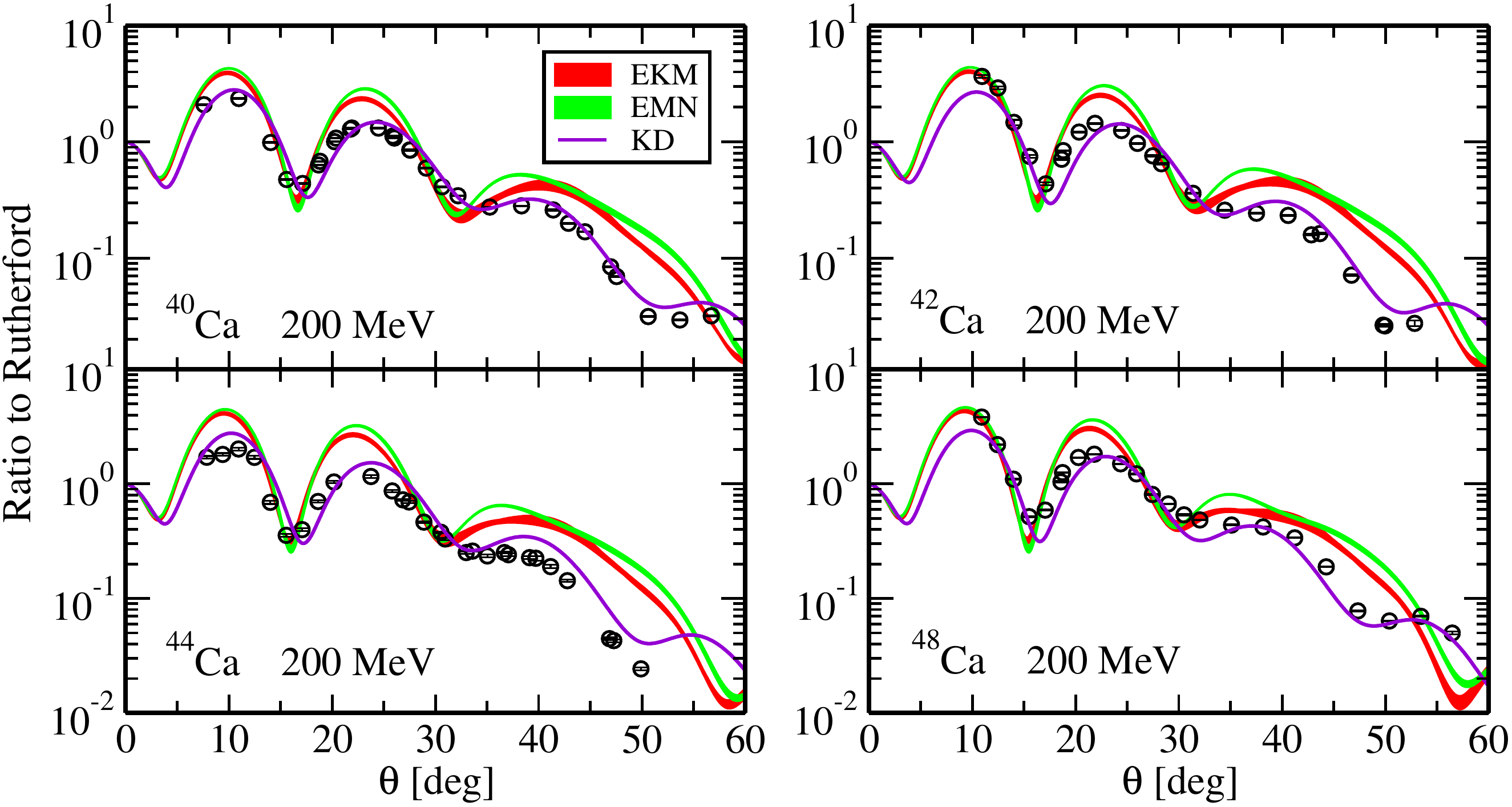}
\caption{ (Color online) The same as in Fig.~\ref{fig_o} for ${}^{40,42,44,48}$Ca isotopes at 200 MeV. 
Experimental data from Refs. \cite{kelly,exfor}.  \label{fig_ca_iso} }
\end{center}
\end{figure}

\begin{figure}[t]
\begin{center}
\includegraphics[scale=0.6]{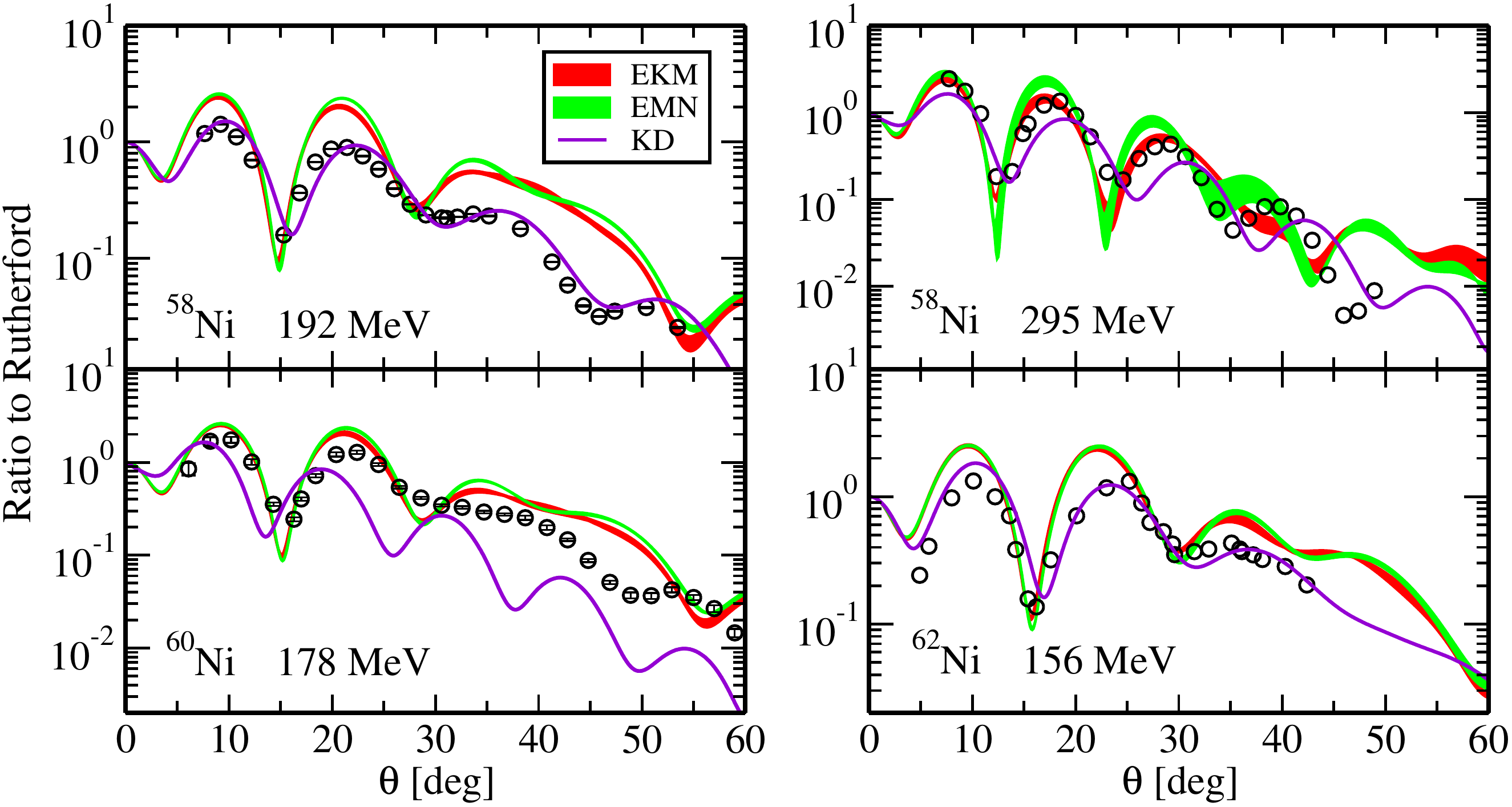}
\caption{ (Color online) The same as in Fig.~\ref{fig_o} for Ni isotopes: 
${}^{58}$Ni at $E = 192$ and $295$ MeV,  ${}^{60}$Ni at $E = 178$ MeV, and  ${}^{62}$Ni 
at $E = 156$ MeV. Experimental data from Refs. \cite{kelly,exfor}.  \label{fig_ni_iso}  }
\end{center}
\end{figure}

\begin{figure}[t]
\begin{center}
\includegraphics[scale=0.6]{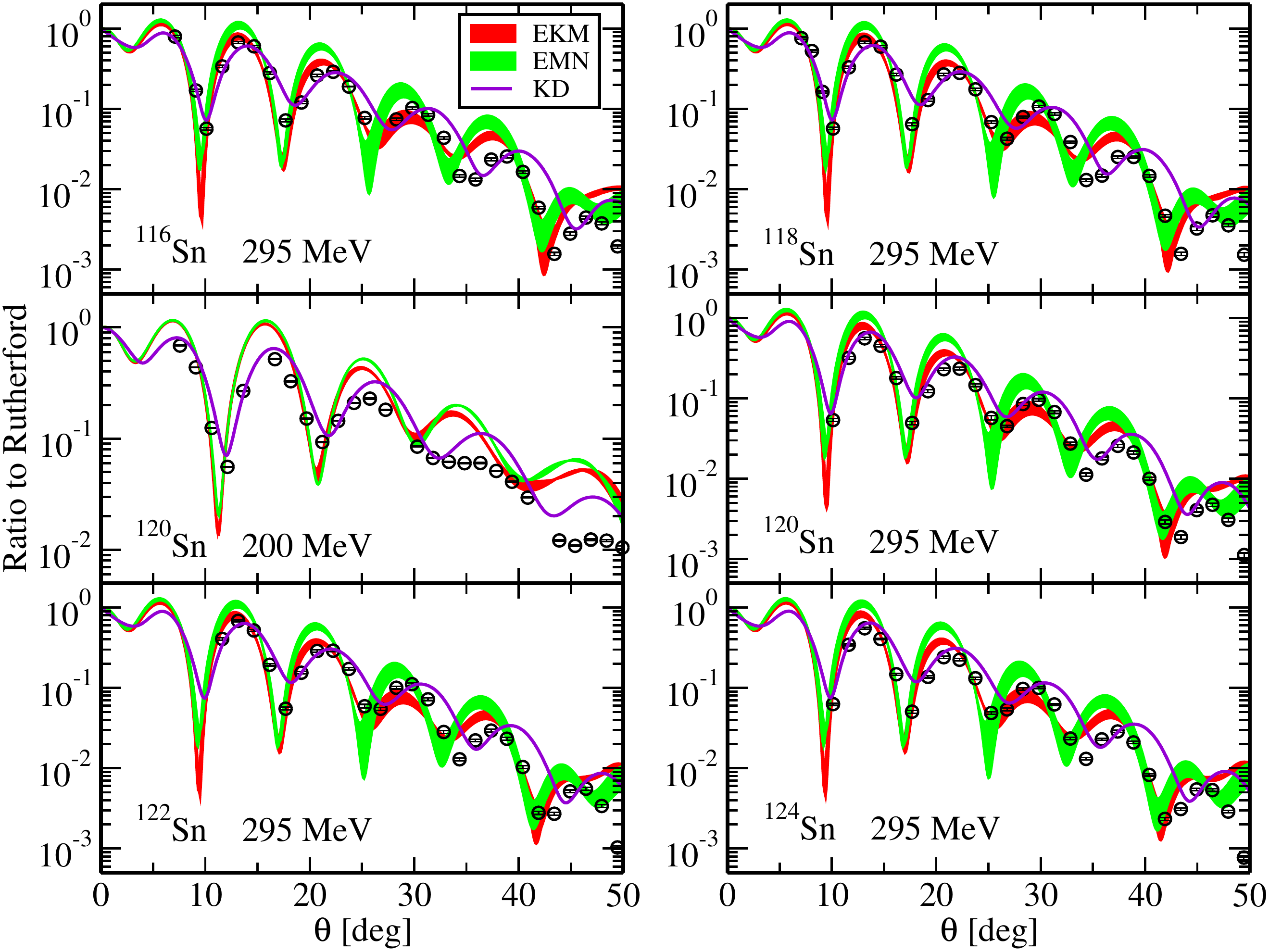}
\caption{(Color online)  The same as in Fig.~\ref{fig_o} for Sn isotopes: 
${}^{120}$Sn at $E = 200$ MeV and ${}^{116,118,120,122,124}$Sn at $E = 295$ MeV. Experimental 
data from Refs. \cite{kelly,exfor}. \label{fig_sn_iso} }
\end{center}
\end{figure}

\begin{figure}[t]
\begin{center}
\includegraphics[scale=0.6]{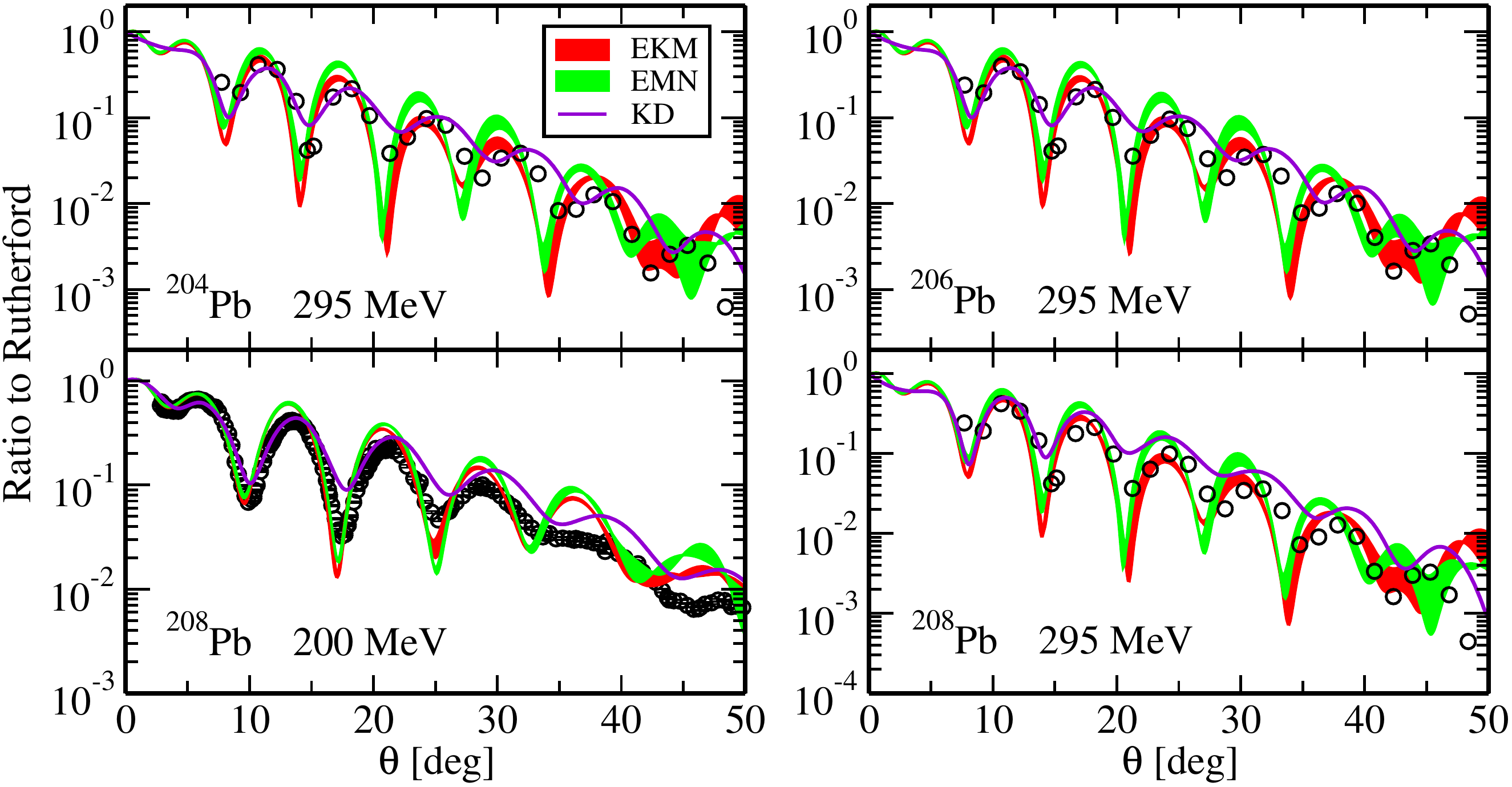}
\caption{(Color online)  The same as in Fig.~\ref{fig_o} for Pb isotopes: ${}^{208}$Pb at $E = 200$ MeV 
and ${}^{204,206,208}$Pb at $E = 295$ MeV. Experimental data from Refs. \cite{kelly,exfor}.  \label{fig_pb_iso} }
\end{center}
\end{figure}

The ratios of the differential cross sections to the Rutherford cross sections for elastic proton scattering off calcium, nichel, tin, and lead isotopes are shown in Figs.~\ref{fig_ca_iso},
\ref{fig_ni_iso}, \ref{fig_sn_iso}, and \ref{fig_pb_iso}. The results are compared with the experimental data taken from Refs. \cite{kelly,exfor}.

All the results for  ${}^{40,42,44,48}$Ca isotopes in Fig.~\ref{fig_ca_iso} are for an incident proton energy of 200 MeV. 
The experimental database used to generate the KD potential includes ${}^{40}$Ca at $E =200$ MeV. In Fig.~\ref{fig_ca_iso} KD 
gives indeed an excellent agreement with ${}^{40}$Ca data, and a good agreement also for the other isotopes. 
The results with the EKM and EMN potentials are very close to each other, the uncertainty bands are narrow, 
and the agreement with data, which is reasonable and of about the same quality for all the isotopes, is however 
somewhat worse than with KD, in particular at larger angles. At lower angles the EKM and EMN results well 
reproduce the behaviour of the experimental cross section, which is sometimes a bit overestimated by the calculations. 
A better agreement with data would presumably be obtained improving or reducing the approximations adopted in the calculation of the microscopic OP.   

In Fig.~\ref{fig_ni_iso} we show the results for ${}^{58}$Ni at $E = 192$ and 295 MeV,  ${}^{60}$Ni at $E = 178$ MeV, and ${}^{62}$Ni  at $E = 156$ MeV.
The experimental database used to generate the KD potential includes  ${}^{58}$Ni up to 200 MeV and ${}^{60}$Ni up to 65 MeV. 
For ${}^{58}$Ni KD gives a good description of the data  at 192 MeV, while a much worse agreement is obtained at the higher energy of 
295 MeV, where only the overall behavior of the experimental cross section is reproduced by the phenomenological OP. 
The EKM and EMN results give a better and reasonable description of the data at 295 MeV, up to $\theta \sim 40^{\circ}$. 
At 192 MeV the microscopic OP can roughly describe the shape of the experimental cross section, 
but the size is somewhat overestimated. KD gives only a poor description of the data for ${}^{60}$Ni at 178 MeV and a very good agreement 
for ${}^{62}$Ni at 156 MeV. The microscopic OP gives a better and reasonable agreement with  the ${}^{60}$Ni data, over all the angular distribution, 
while for ${}^{62}$Ni the results are a bit larger than those of the KD potential. 
The EKM and EMN results are always very close to each other and the bands are generally narrow.    

The results for ${}^{116,118,120,122,124}$Sn isotopes at 295 MeV and for ${}^{120}$Sn at 200 MeV are 
displayed in Fig.~\ref{fig_sn_iso}. In this case all the OPs give qualitatively similar results and 
a reasonable agreement with data, in particular, for $\theta \le 20^{\circ}$. The agreement generally declines 
for larger angles. KD gives a better description of ${}^{120}$Sn data at 200 MeV, where the EKM and EMN 
results are a bit larger than the data at the maxima and a bit lower at the minima. We note that ${}^{120}$Sn 
is included in the experimental database for the KD potential for proton energies up to 160 MeV. At 295 MeV, the 
microscopic OP gives, in general, a slightly better agreement with the data than KD for all the tin isotopes shown in the figure.

The results for ${}^{204,206,208}$Pb isotopes at 295 MeV and for ${}^{208}$Pb data at 200 MeV 
are displayed in Fig.~\ref{fig_pb_iso}. Also in this case the experimental cross section at 200 MeV 
is well described by KD, the agreement is better than with the microscopic OP. The experimental 
database for KD includes ${}^{208}$Pb up to 200 MeV. At 295 MeV a better agreement with data is 
generally given by the EKM and EMN results, in particular by EMN: for all the three isotopes 
considered, the two results practically overlap for $\theta \le 20^{\circ}$, where they are also very 
close to the KD result, then they start to separate and the EMN result is a bit larger than the 
EKM one and in better agreement with data. We point out that the uncertainty bands, that are 
generally narrow, in this case become larger increasing the scattering angle, when also the agreement with data declines. 

\begin{figure}[t]
\begin{center}
\includegraphics[scale=0.6]{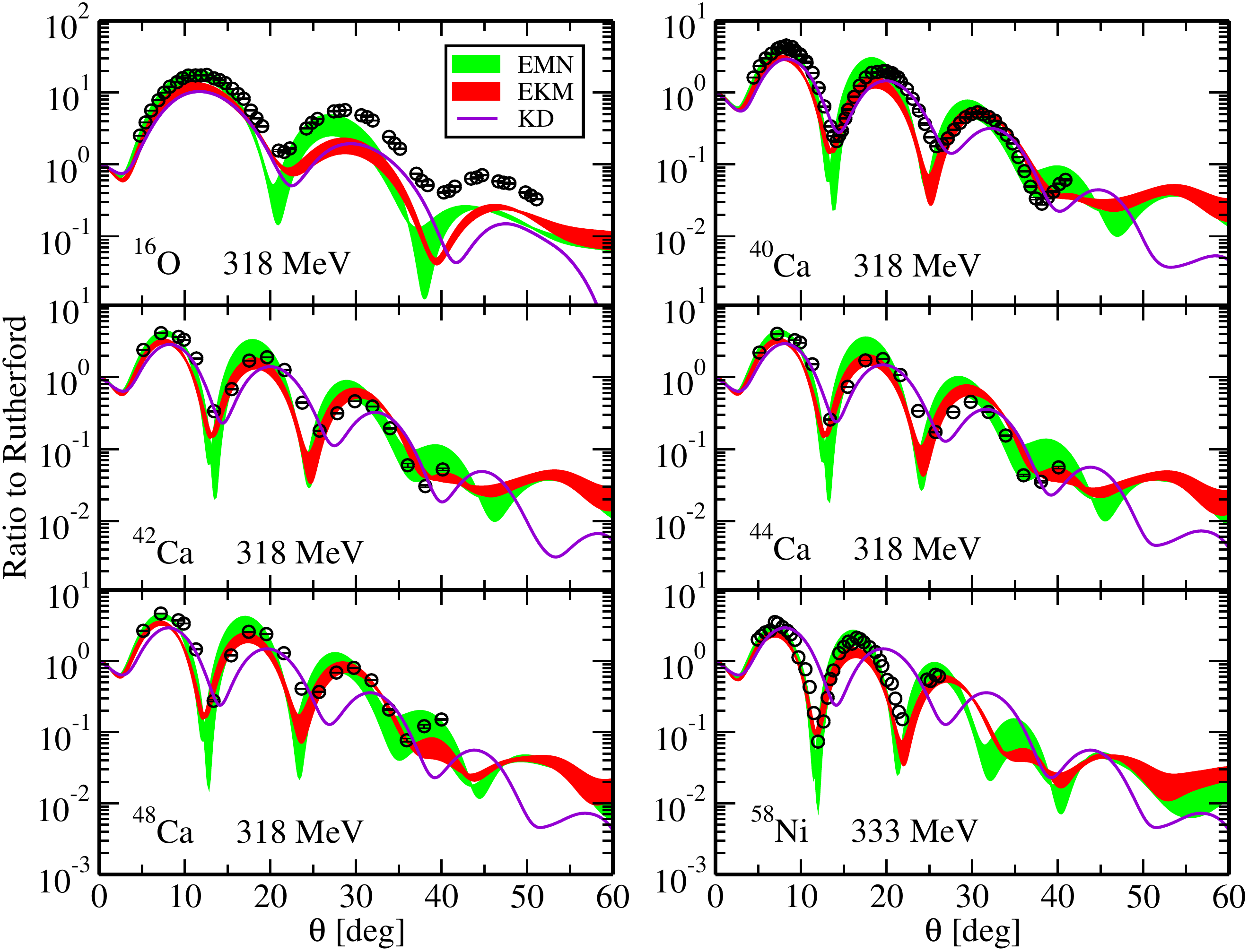}
\caption{(Color online)  The same as in Fig.~\ref{fig_o} for ${}^{16}$O and ${}^{40,42,44,48}$Ca 
at $E = 318$ MeV and ${}^{58}$Ni at $E = 333$ MeV.  Experimental data from Refs. \cite{kelly,exfor}.  \label{fig_high} }
\end{center}
\end{figure}

The results that we have shown till now indicate that, in comparison with the phenomenological KD potential, 
our microscopic OP, in spite of the approximations made to derived it, has a comparable and 
in some cases even better predictive power in the description of the cross sections on 
the isotopic chains and energy range here considered. KD is able to give a better and excellent 
description of data in specific situations, in particular, in the case of nuclei included in the experimental database 
used to generate the original KD potential and at the lower energies considered. 
For energies above 200 MeV our microscopic OP gives, in general, a better agreement with data. 
This conclusion is confirmed by the results shown in Fig.~\ref{fig_high}, where the ratios of 
the differential cross sections to the Rutherford cross sections are displayed for elastic 
proton scattering off ${}^{16}$O and ${}^{40,42,44,48}$Ca at $E = 318$ MeV and ${}^{58}$Ni at $E = 333$ MeV 
in comparison with the data taken from Refs. \cite{kelly,exfor}. The differences between the phenomenological 
and microscopic OPs increase with the increasing scattering angle and proton energy. For ${}^{58}$Ni at 333 MeV 
both EKM and EMN give a much better and very good description of data. In the other cases KD is able to describe 
data only at the lowest angles. The EKM and EMN results are in general very close to each other. In both cases the width of the uncertainty 
bands increases at larger scattering angles but the uncertainties are not crucial for the comparison with data.

\begin{figure}[t]
\begin{center}
\includegraphics[scale=0.6]{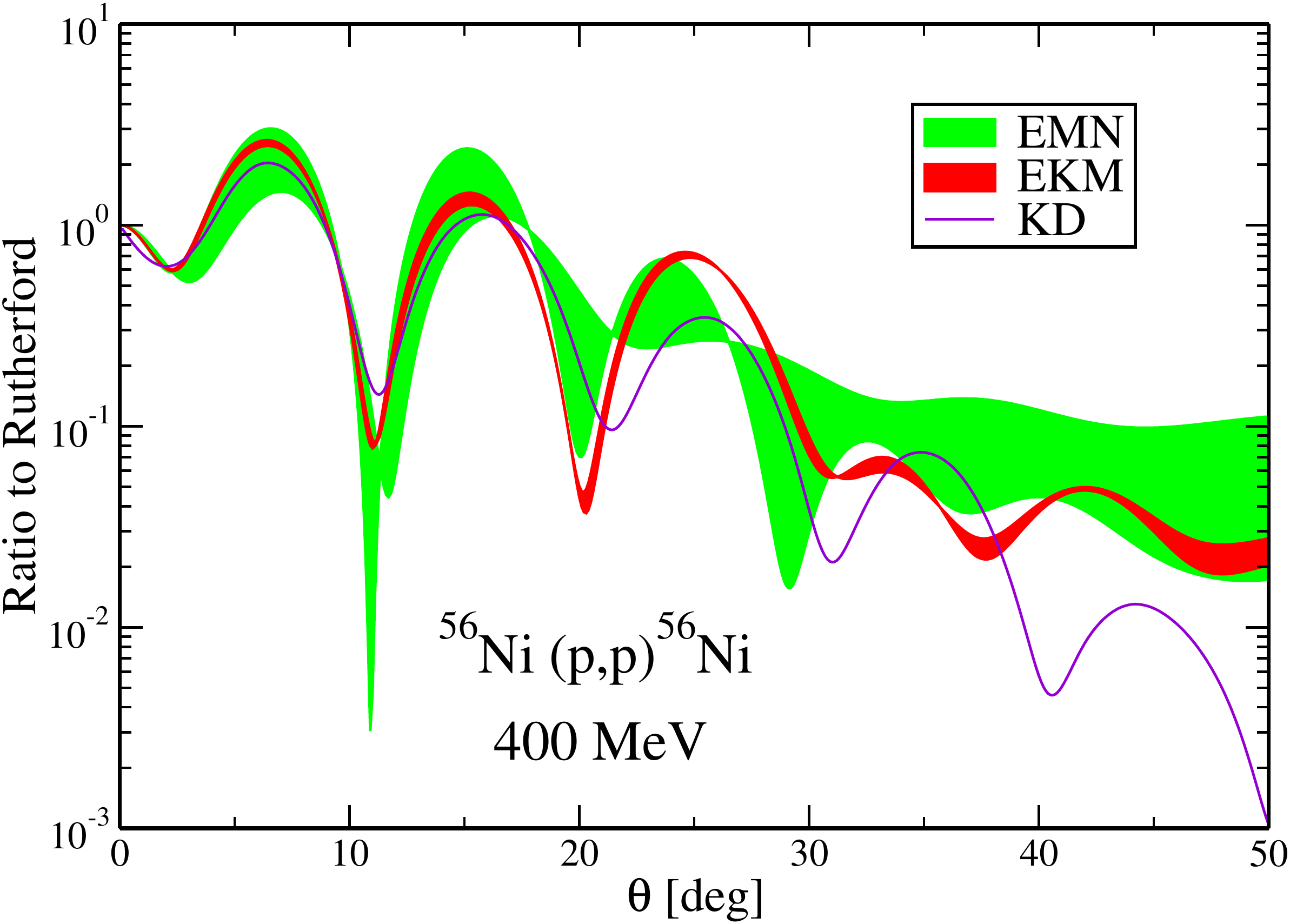}
\caption{(Color online)  The same as in Fig.~\ref{fig_o} for ${}^{56}$Ni at $E = 400$ MeV.
\label{fig_exl}}
\end{center}
\end{figure}

As mentioned in the Introduction (Sec.~\ref{sec_intro}), in the near future new experimental data will be available for exotic nuclei \cite{fair, exl}. For this purpose, in Fig. \ref{fig_exl}
we show theoretical predictions for EKM and EMN potentials for the test case $^{56}$Ni(p,p)$^{56}$Ni at an energy of $400$ MeV in inverse kinematics. Even if the energy scale involved is beyond
the supposed range of validity of our approach, it is interesting to see if microscopic potentials and KD look reasonable and if error bands are acceptable. We only included a
selection of the potentials because of the energy: $R=0.8$ and $0.9$ fm for EKM potentials and $\Lambda = 500$ and $550$ MeV for the EMN ones. At 400 MeV KD and EKM
give similar predictions with reasonable shapes as a function of the angle $\theta$. The EMN potentials seem under control only for small angles ($\theta \le 30^{\circ}$).
Unfortunately, experimental data are still under scrutiny and not yet published \cite{datani56}. From this figure we can thus conclude that, contrary to the EMN potentials,  the EKM potentials have not yet reached the limit beyond which the chiral expansion scheme breaks down. Of course, this limit is not unique and depends on the regularization scheme
adopted to derive the NN interaction.

\begin{figure}[t]
\begin{center}
\includegraphics[scale=0.6]{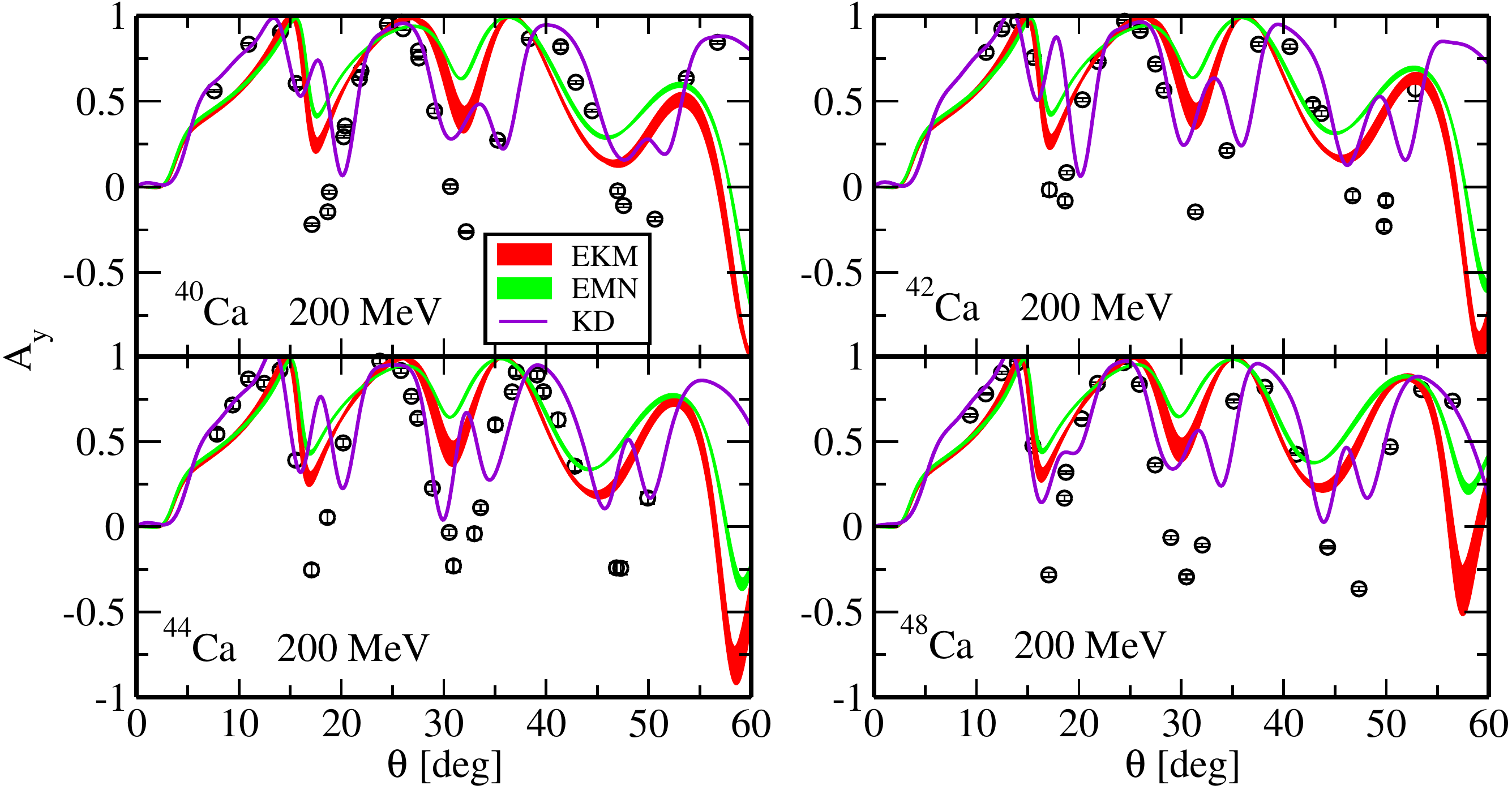}
\caption{(Color online) The same as in Fig.~\ref{fig_ca_iso} but for the analyzing power $A_y$. 
Experimental data from Refs. \cite{kelly,exfor}. \label{fig_ca_ay}  }
\end{center}
\end{figure}
\begin{figure}[t]
\begin{center}
\includegraphics[scale=0.6]{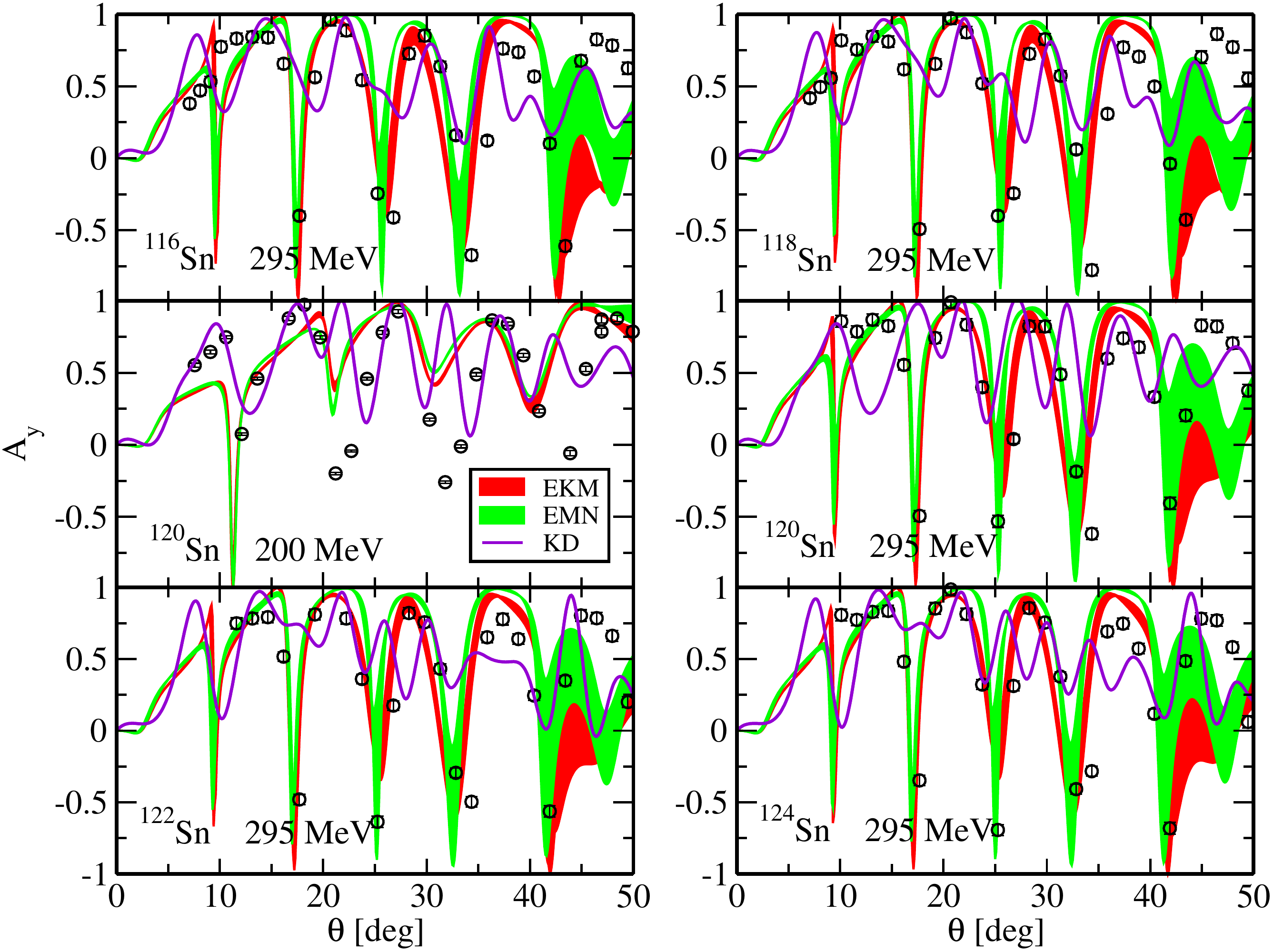}
\caption{(Color online) The same as in Fig.~\ref{fig_sn_iso} but for the analyzing power $A_y$. 
Experimental data from Refs. \cite{kelly,exfor}.   \label{fig_sn_ay} }
\end{center}
\end{figure}
\begin{figure}[t]
\begin{center}
\includegraphics[scale=0.6]{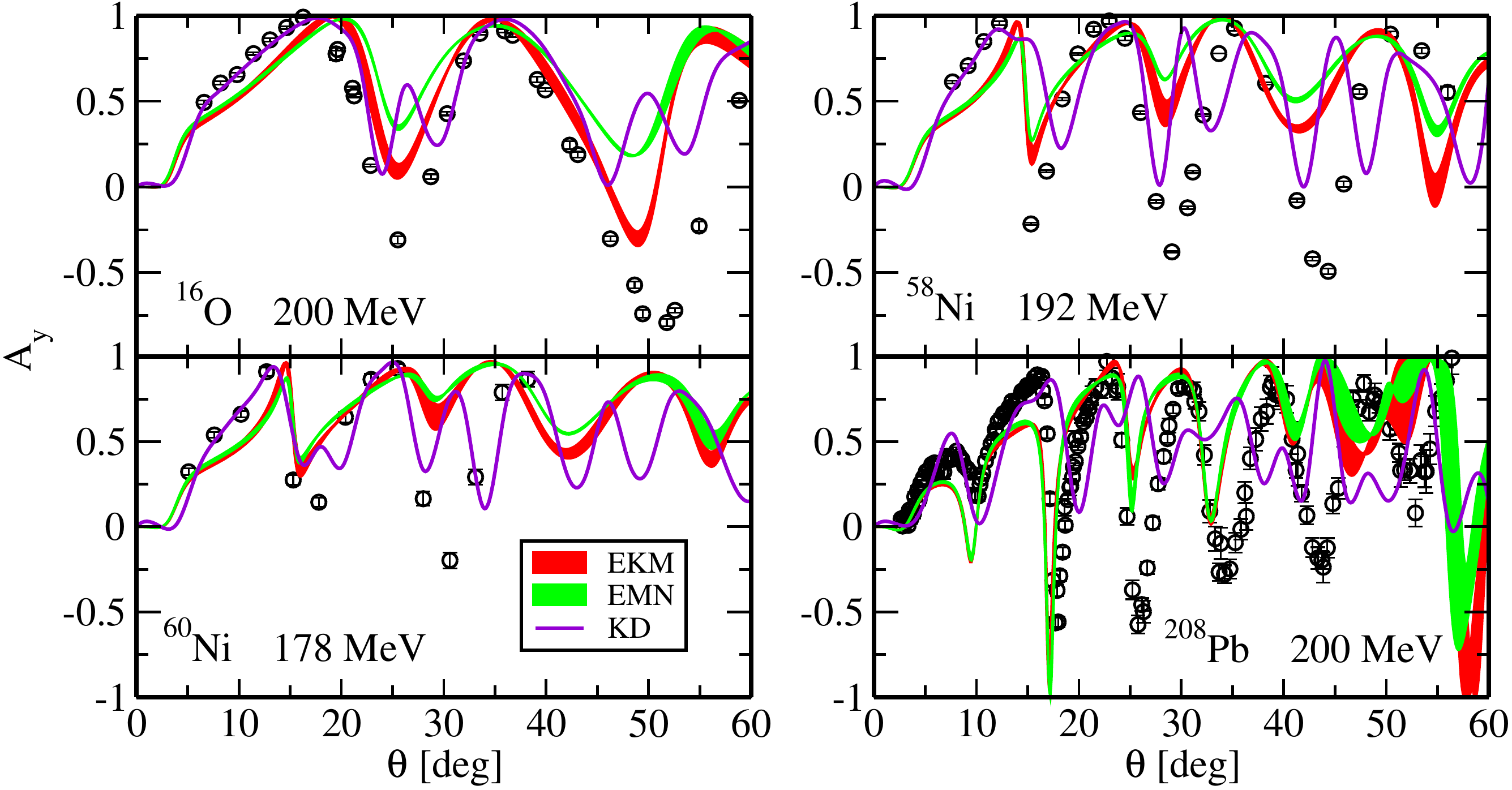}
\caption{(Color online) Analyzing power $A_y$ as a function of the angle $\theta$ for elastic proton scattering 
on ${}^{16}$O, and ${}^{208}$Pb at $E = 200$ MeV,  ${}^{58}$Ni at $E = 192$ MeV, and ${}^{60}$Ni at $E = 178$ MeV. 
Experimental data from Refs. \cite{kelly,exfor}.  \label{fig_onipb_ay}  }
\end{center}
\end{figure}

In Figs.~\ref{fig_ca_ay}, \ref{fig_sn_ay}, and \ref{fig_onipb_ay} we show the analyzing power $A_y$ for some 
of the same nuclei and at the same proton energies presented in Figs.~\ref{fig_o}, \ref{fig_ca_iso}, \ref{fig_ni_iso}, 
\ref{fig_sn_iso}, and \ref{fig_pb_iso}. 

The analyzing power for calcium isotopes is shown Fig.~\ref{fig_ca_ay}, which corresponds to Fig.~\ref{fig_ca_iso} 
for the ratio of the differential cross section to the Rutherford cross section. Polarization observables are 
usually more difficult to reproduce and also in this case the agreement with data is far from perfect,  but 
all the results are able to describe the overall behavior of the experimental $A_y$, in particular at lower angles. 
A better result in comparison with data is given in this case by the phenomenological KD potential.
The differences between the EKM and EMN results are small, but the error bands get large at the largest angles considered. 

The results for tin isotopes in Fig.~\ref{fig_sn_ay} correspond to the results of Fig.~\ref{fig_sn_iso} for the 
ratio ${\frac{{\rm d}\sigma}{{\rm d}\Omega}}/{\frac{{\rm d}\sigma}{{{\rm d}\Omega}}}_{\rm Ruth}$. 
Also in this case the agreement with data is worse than in Fig.~\ref{fig_sn_iso} and it is difficult to judge which 
OP gives the better description of the experimental $A_y$: KD is somewhat 
better at 200 MeV and the microscopic OP at 295 MeV. 
It must be emphasized that the extension to 1 GeV was performed to have the total reaction cross section under control up to this energy with no concerns about polarization observables. The bad performance in the description of $A_y$ is anyhow not surprising if we consider that Woods-Saxon like form factors are not supposed to work properly above 200 MeV \cite{koning}.

The analyzing power for ${}^{16}$O and ${}^{208}$Pb at 200 MeV, ${}^{58}$Ni at 192 MeV, and ${}^{60}$Ni at 178 
MeV are displayed in Fig.~\ref{fig_onipb_ay}. In the case of ${}^{16}$O, the results basically confirm what was 
already found for the ratio in Fig.~\ref{fig_o}: KD gives a good description of data for lower angles, 
in particular  for $\theta \le 20^{\circ}$, EKM and EMN give a less good description of data at lower angles 
but a reasonable agreement  up to $\theta \sim 50^{\circ}$. The experimental analyzing powers of ${}^{58}$Ni 
and ${}^{60}$Ni are well described by KD at the lowest angles, but over all the angular distribution the 
general agreement (or disagreement) with data of the microscopic and phenomenological OPs is of about the same quality. 
Also in the case of ${}^{208}$Pb KD describes the experimental $A_y$ well for $\theta \le 20^{\circ}$, better 
than the EKM and EMN results, which, on the other hand, are in a somewhat better (although not perfect) 
agreement with data for larger angles.


\section{Conclusions}
\label{sec_concl}

In recent papers~\cite{Vorabbi:2015nra,Vorabbi:2017rvk} we derived a microscopic optical potential for elastic pA scattering from NN chiral potentials at fourth
order (N$^3$LO) and fifth order (N$^4$LO), with the purpose to study the domain of applicability of NN chiral potentials to the construction of an 
optical potential, to investigate convergence patterns,  and to assess the theoretical errors associated with the truncation of the chiral expansion.
Numerical examples for the cross section and polarization observables of elastic proton scattering on $^{12}$C, $^{16}$O, and $^{40}$Ca nuclei were presented and compared with available experimental data.
Our results indicated that building an optical potential within the chiral perturbation theory is a promising approach for describing elastic proton-nucleus  scattering and they allowed us to conclude that convergence has satisfactorily been achieved at N$^4$LO.


In the present work we have extended our previous investigation 
to isotopic chains exploring the mass number dependence and the energy range of applicability of our microscopic optical potential.
As a benchmark, we have tested our calculations against one of the best phenomenological parametrizations developed by Koning and Delaroche \cite{Koning:2003zz} 
and, of course, experimental data where available.

Our main goal was to check the robustness of our approach and the capability of our optical potential to be applied to exotic nuclei, {\it i.e.} nuclei with values of proton-to-neutron ratio far from the stability, where  phenomenological 
models might be unreliable.

Numerical results have been presented for the unpolarized differential cross section 
and the analyzing power 
of elastic proton scattering off calcium, nickel, tin, and lead isotopes in a proton energy range between 156 and 333 MeV.
A theoretical prediction for the cross section of elastic proton scattering off  $^{56}$Ni at 400 MeV has also been presented, which is of interest for the future EXL experiment on exotic nuclei at FAIR \cite{fair,exl}, although the energy is beyond the supposed range of validity of the chiral potentials.

Because of the renormalization procedure, NN chiral potentials with almost the same
level of accuracy (i.e. the $\chi^2$ associated to the reproduction of NN phase shifts) come with different values
of the cut-off parameters. We have restricted our calculations to a limited set of the aforementioned parameters and we have associated theoretical error bands
to the uncertainties produced by the different parameters.

The agreement of our present results with empirical data is sometimes worse and sometimes better but overall comparable to the agreement given by the   phenomenological OP, in particular  for energies close to 200 MeV and above 200 MeV. 
The example shown at 400 MeV suggests that at this energy the EKM potentials have not yet reached the limit after which the chiral expansion scheme breaks down.  

The microscopic optical potential generally provides a qualitatively similar  agreement with data for all the nuclei of an isotopic chain. This clearly shows that changing the values of $A$ does not affect the predictive power of our optical potential. 
The agreement is  worse for the analyzing power than for the cross section and in general declines for larger values of the scattering angle. 

A better description of empirical data requires a more sophisticated model for the microscopic optical potential. 
Work is in progress about three major improvements: calculation of the full-folding integral (to go beyond the optimum factorization),
treatment of nuclear-medium effects (to go beyond the impulse approximation), and the inclusion of three-body forces.





\section{Acknowledgements}
The authors are deeply 
grateful to E. Epelbaum (Institut f\"ur Theoretische Physik II,
Ruhr-Universit\"at Bochum, Germany) for providing the chiral potential of Ref. \cite{Epelbaum:2014sza,Epelbaum:2014efa} and R. Machleidt (Department of Physics, University
of Moscow, Idaho, USA) for the chiral potential of Ref. \cite{Entem:2014msa,Entem:2017gor}. We also would like to acknowledge useful communications with A. Koning (Nuclear Research and Consultancy Group NRG, Petten, The Netherlands) and T. Kr\"oll (Institut f\"ur Kernphysik, Technische Universit\"at Darmstadt, Germany).

\bibliography{biblio}


\newpage

\end{document}